\RequirePackage{lineno}
\documentclass[prd,twocolumn,nofootinbib,superscriptaddress]{revtex4}
\usepackage{cancel}
\usepackage{graphicx}
\usepackage{epsfig}
\usepackage{amsmath,amsfonts,amssymb}
\usepackage{t1enc}

\newcommand{\mnmw}{(M_{W_R},m_N)}

\begin{document}


\title{Flavour in heavy neutrino searches at the LHC}

\author{J. A. Aguilar-Saavedra}
\email{jaas@ugr.es}
\affiliation{Departamento de F\'{\i}sica Te\'orica y del Cosmos, Universidad de Granada,\\
 E-18071 Granada, Spain}
\author{F. Deppisch}
\email{f.deppisch@ucl.ac.uk}
\affiliation{Department of Physics and Astronomy, University College London, \\
London WC1E 6BT, United Kingdom}
\author{O. Kittel}
\email{kittel@th.physik.uni-bonn.de}
\affiliation{Departamento de F\'{\i}sica Te\'orica y del Cosmos, Universidad de Granada,\\
 E-18071 Granada, Spain}
\author{J. W. F. Valle}
\email{valle@ific.uv.es}
 \affiliation{AHEP Group, Institut de F\'{\i}sica Corpuscular --
  C.S.I.C./Universitat de Val{\`e}ncia \\
  Edificio Institutos de Paterna, Apt 22085, E--46071 Valencia, Spain}


\begin{abstract}
Heavy neutrinos at the TeV scale have been searched for at the LHC in the context of left-right models, under the assumption that they couple to the electron, the muon, or both. We show that current searches are also sensitive to heavy neutrinos coupling predominantly to the tau lepton, and present limits can significantly constrain the parameter space of general flavour mixing. 
\end{abstract}

\maketitle

The discovery of neutrino oscillations~\cite{Nakamura:2010zzi}
demonstrates that lepton flavour is not conserved in nature, providing
evidence for neutrino mass and new physics~\cite{Schwetz:2008er}.
Seesaw schemes implemented in a left-right
context~\cite{Minkowski:1977sc,Mohapatra:1979ia} not only provide a natural
origin for neutrino mass via the introduction of three heavy right-handed neutrinos $N_i$, $i=1-3$, but also open the possibility of their direct observation at high energies. In the absence of extra gauge bosons or scalars, the production cross sections of heavy neutrinos with masses $m_N \gtrsim 100$ GeV at the Large Electron Positron (LEP) collider~\cite{Dittmar:1989yg}, HERA~\cite{Buchmuller:1990vh} and hadron colliders~\cite{Datta:1993nm} are rather small, due to the small coupling of heavy neutrinos to SM particles~\cite{Nardi:1994iv,Tommasini:1995ii,delAguila:2008pw} (see~\cite{del Aguila:2006dx} for a review and additional references).
In contrast, in models with left-right symmetry heavy neutrinos have right-handed interactions with charged leptons $\ell = e,\mu,\tau$, 
\begin{equation}
\mathcal{L} = -\frac{g_R}{\sqrt 2} V_{\ell N_i}^R \bar \ell_R \gamma^\mu N_{iR} W_{R\mu}^{-} + \text{H.c.} \,,
\end{equation}
with mixing matrix elements $V_{\ell N_i}^R$ that can be of order unity. Consequently, they can be produced singly at the Large Hadron Collider (LHC), in
\begin{equation}
pp \to W_R \to \ell N_i \,,
\label{ec:prod}
\end{equation}
with a relatively large cross section, provided the $W_R$ mass $M_{W_R}$ is larger than $m_N$, so that the $s$-channel $W_R$ boson is on its mass shell. The heavy neutrinos can then decay via an off-shell $W_R$ boson,\footnote{If the mixing of heavy neutrinos with light leptons is close to the upper experimental limits, the decays $N \to \ell W, \nu Z,\nu H$ can be relevant in some regions of the $(m_N,M_{W_R})$ parameter space~\cite{delAguila:2009bb}.} 
\begin{equation}
N_i \to \ell' jj\,,~\ell' tb \,. 
\label{ec:dec}
\end{equation}
The resulting signal of two charged leptons plus two jets has already been proposed in the literature as a golden channel for the detection of heavy neutrinos and new $W_R$ bosons~\cite{Keung:1983uu} and, possibly, to shed light upon the neutrino mass generation mechanism, for which there is no clue yet. Indeed, this is a much more favourable scenario than without extra $W_R$ bosons, where the heavy neutrino production is overwhelmed by the SM background~\cite{delAguila:2007em}, and the discovery potential is also better than in models with an extra $Z'$ boson~\cite{delAguila:2007ua,AguilarSaavedra:2009ik}, where heavy neutrinos are produced in pairs.

Previous sensitivity estimations~\cite{Ferrari:2000sp,Gninenko:2006br} for the process in Eqs.~(\ref{ec:prod}),(\ref{ec:dec}) and preliminary searches \cite{CMSN} have taken as benchmark scenarios for the study the case of heavy neutrinos $N_e,N_\mu$ coupling to the electron or the muon only. This has also been the case for early reinterpretations of LHC searches~\cite{Nemevsek:2011hz}. Recently, the ATLAS Collaboration has gone a step further~\cite{atlas} and has searched for neutrinos $N_{1,2}$ mixing  maximally with the electron {\it and} the muon, $|V_{eN_i}^R| = |V_{\mu N_i}^R| = 1/\sqrt 2$, $V_{\tau N_i}^R = 0$ with $i=1,2$. However, these approximations may be too simplistic to describe TeV-scale seesaw, if it is realised in nature. Indeed, a rich flavour structure appears in the measured light neutrino mixing with charged leptons, with the atmospheric angle being close to maximal, the solar angle also large~\cite{Schwetz:2008er}, and the third mixing angle $\theta_{13} \approx 8.8 \pm 0.8^\circ$ significantly different from zero as recently discovered~\cite{An:2012eh}.  Motivated by this consideration, in this paper we show that the ATLAS search in Ref.~\cite{atlas} actually provides stringent constraints on heavy neutrinos with predominant mixing with the tau, $V_{\tau N_i}^R \gg V_{eN_i}^R, V_{\mu N_i}^R$. The key for this observation is that no requirement is made in this search regarding the missing energy of the events. Thus, the ATLAS search~\cite{atlas} is sensitive to the process in Eqs.~(\ref{ec:prod}),(\ref{ec:dec}) with $\ell,\ell' = \tau$ and subsequent $\tau$ leptonic decay, where the main penalty with respect to the channels with $\ell,\ell' = e,\mu$ is the $\tau$ leptonic branching ratio of around $1/3$. The approach to the study of heavy neutrino production presented here is complementary to previous ones. Instead of fixing the flavour structure to study the mass reach in terms of $M_{W_R}$ and $m_N$, since we want to concentrate on flavour, we follow the approach in~\cite{del Aguila:2005mf} and fix $\mnmw$ to study the size of the resulting signals in terms of the flavour structure of the heavy neutrino couplings. Such general approach would be unavoidable in case that a positive signal were observed. 

\begin{figure*}[htb]
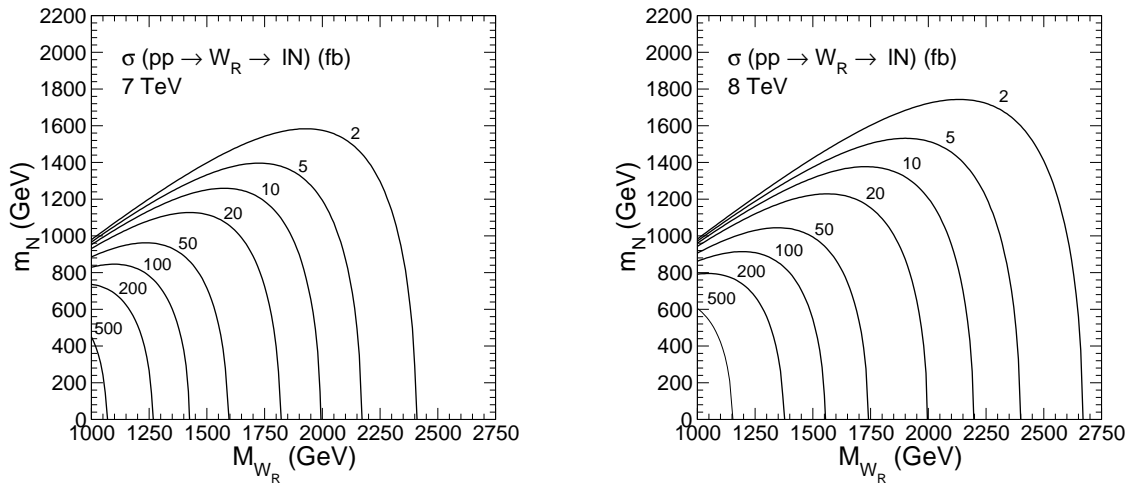

\begin{center}
\begin{tabular}{ccc}
\epsfig{file=fig1a.eps,width=6.8cm,clip=} & \quad\quad\quad\quad &
\epsfig{file=fig1b.eps,width=6.8cm,clip=}
\end{tabular}
\caption{Cross sections for the production of a single heavy neutrino at CM energies of 7 TeV (left) and 8 TeV (right).}
\label{fig:xsec}
\end{center}
\end{figure*}

For simplicity, we assume that only one heavy neutrino $N \equiv N_1$ is lighter than $W_R$. (In case there are more than one, their dilepton signals peak at different $\ell' jj$ invariant masses and interference effects are negligible due to their small heavy neutrino intrinsic width.) In our calculations we take $g_R$ equal to the SM weak coupling $g_W$, and the right-handed quark mixing matrix as diagonal. Relaxing the latter assumptions would only amount to an overall scale factor in our cross sections, which does not affect our conclusions. To a good approximation we can neglect the left-right mixing in the leptonic charged current, so that $|V_{eN}^R|^2 + |V_{\mu N}^R|^2 + |V_{\tau N}^R|^2 = 1$ by unitarity~\cite{Schechter:1980gr}.
The production cross sections corresponding to this heavy neutrino, summing over $\ell=e,\mu,\tau$, are presented in Fig.~\ref{fig:xsec} for centre of mass (CM) energies of 7 and 8 TeV. Notice that these cross sections are independent of the heavy neutrino mixing due to the unitarity constraint. Higher-order corrections increase these cross sections by a factor $K\sim 1.3$ for the range of $W_R$ masses considered~\cite{Gavin:2010az}.

In order to estimate the sensitivity to general flavour mixing, we have performed a fast simulation analysis of the process in Eqs.~(\ref{ec:prod}),(\ref{ec:dec})  using {\sc Triada}~\cite{AguilarSaavedra:2009ik,delAguila:2008cj} for the signal generation, {\sc Pythia}~\cite{Sjostrand:2006za} for hadronisation and {\sc AcerDET}~\cite{RichterWas:2002ch} for the simulation of a generic LHC detector. We have selected two benchmark points in the $\mnmw$ parameter space for illustration. The first one, $M_{W_R} = 1.5$ TeV, $m_N = 0.8$ TeV, with a heavy neutrino production cross section of 44 fb, is allowed by direct searches if $N$ couples either to the electron or the muon. These values of $M_{W_R}$ and $m_N$ coincide with a benchmark point in~\cite{atlas} where, on the other hand, results are given for {\it two} heavy neutrinos with their contributions summed (we only consider one heavy neutrino $N=N_1$ here.) For our second benchmark we choose slightly ligher $W_R$ and $N$, $M_{W_R} = 1.2$ TeV, $m_N = 0.6$ TeV, corresponding to a signal cross section of 179 fb. These masses are excluded by direct searches if $N$ does not couple to the tau, but are allowed otherwise. Moreover, the value taken for the $W_R$ mass is disfavoured by indirect limits $M_{W_R} \gtrsim 1.6$ TeV from the $K_L-K_S$ mass difference~\cite{Barenboim:1996nd}. Still, we have chosen this second scenario to illustrate the type of flavour constraints that will be achieved for higher masses and larger luminosity, also with the CM energy upgrade at 8 TeV that approximately doubles the cross sections (see Fig.~\ref{fig:xsec}). Here it is worthwhile remarking that in a large part of the $\mnmw$ parameter space the sensitivity to heavy neutrino production with arbitrary flavour mixing corresponds to a simple rescaling of the results presented here by the $\ell N$ production cross section, according to Fig.~\ref{fig:xsec}. The exception is for the region with $m_N \ll M_{W_R}$, where the heavy neutrino from $W_R$ decay is highly boosted and the charged lepton in $N \to \ell jj$ overlaps with the hadronic jets, making detection efficiencies decrease.

For our two selected benchmark scenarios we have generated nine signal samples of 50000 events, each sample corresponding to a flavour combination $\ell,\ell' = e,\mu,\tau$. 
On these simulated samples we have mimicked the event selection criteria in Ref.~\cite{atlas}, namely
\begin{itemize}
\item Exactly two leptons with transverse momentum $p_T > 25$ GeV, and pseudo-rapidity $|\eta| < 2.47$ for electrons, $|\eta| < 2.4$ for muons. Electrons in the range $1.37 < |\eta| < 1.52$ are excluded.
\item At least one jet with $p_T > 20$ GeV and $|\eta| < 2.8$.
\item The invariant mass of the two leptons $m_{\ell \ell}$ must be larger than 110 GeV.
\item The sum of transverse momenta of the two leptons and highest $p_T$ jets (including up to two jets) must be larger than 400 GeV.
\item The $W_R$ reconstructed mass must be larger than 400 GeV. For events with at least two jets, this mass is defined as the invariant mass of the two leptons and the two leading jets, $M_{W_R}^\text{rec} = m_{\ell \ell jj}$. For events with only one jet, it includes only this jet, $M_{W_R}^\text{rec} = m_{\ell \ell j}$.
\end{itemize}
The resulting efficiencies for event selection are collected in Table~\ref{tab:eff}.
\begin{table}[htb]
\caption{Efficiencies for final event selection in the two benchmark scenarios characterised by $\mnmw$, for the nine samples with $\ell,\ell'=e,\mu,\tau$.\label{tab:eff}}
\begin{tabular}{lccccccc}
\hline
\hline
& \multicolumn{3}{c}{$(1.5,0.8)$ TeV} && \multicolumn{3}{c}{$(1.2,0.6)$ TeV} \\
$\ell$ $\backslash$ $\ell'$ & $e$ & $\mu$ & $\tau$ & \quad \quad & $e$ & $\mu$ & $\tau$ \\
$e$    & 0.555 & 0.501 & 0.175 
      && 0.552 & 0.499 & 0.157 \\
$\mu$  & 0.524 & 0.480 & 0.172 
      && 0.534 & 0.482 & 0.154 \\
$\tau$ & 0.203 & 0.192 & 0.058 
      && 0.186 & 0.176 & 0.045 \\
\hline
\hline
\end{tabular}
\end{table}
These numbers are inclusive, {\em i.e.} we have summed events with same-sign and opposite-sign lepton pairs, including $ee$, $\mu \mu$ and $e \mu$. (The efficiencies for the exclusive channels are not presented here for brevity.)
For $\ell,\ell'=e,\mu$ we can see that the efficiencies are within the range 40-65\% reported in Ref.~\cite{atlas}, giving confidence that our crude simulation gives results not far from a more realistic one. Interestingly, we observe that for tau leptons the efficiencies are close to the ones corresponding to a mere branching ratio scaling $\text{Br}(\tau \to e/\mu \, \nu \bar \nu) \simeq 1/3$. This is due to the high transverse momentum with which the tau leptons are produced, which is inherited by the secondary electrons and muons from their decay, so that the event selection criteria are easily passed. Moreover, as we have emphasised before, the key to retain these efficiencies for $\ell,\ell' = \tau$ is to drop any requirement on small missing energy, as these events involve up to four neutrinos in the final state. Fortunately, this has been the case in the recent ATLAS search, a fact that allows us to explore the constraints on scenarios with non-zero $\tau$ mixing.

\begin{figure*}[htb]
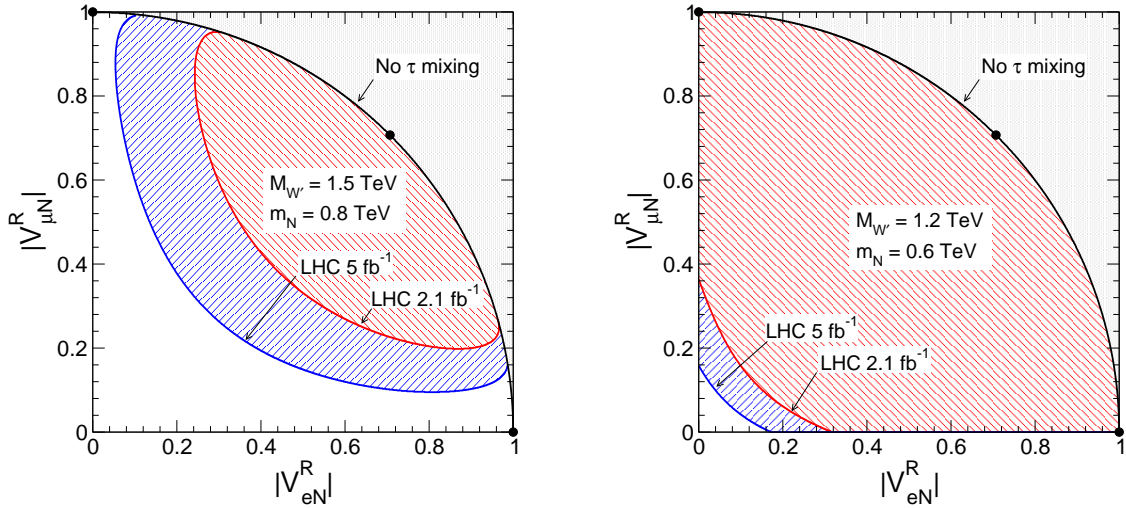

\begin{center}
\begin{tabular}{ccc}
\epsfig{file=fig2a.eps,width=6.8cm,clip=} & \quad\quad\quad\quad &
\epsfig{file=fig2b.eps,width=6.8cm,clip=}
\end{tabular}
\caption{Constraints on arbitrary right handed neutrino mixing parameters in the two benchmark scenarios considered. The gray regions in the upper right side of each plot are forbidden by the unitarity constraint $|V_{eN}^R|^2 + |V_{\mu N}^R|^2 + |V_{\tau N}^R|^2 = 1$.}
\label{fig:lim}
\end{center}
\end{figure*}

Once equipped with the selection efficiencies for each exclusive channel, we can easily compute the number of events for each $\mnmw$ and arbitrary mixing $V_{eN}^R$, $V_{\mu N}^R$, $V_{\tau N}^R$, with the only condition that $|V_{eN}^R|^2 + |V_{\mu N}^R|^2 + |V_{\tau N}^R|^2 = 1$, by an appropriate weighting of the nine samples.
We use the SM background prediction and number of observed events in Ref.~\cite{atlas} for the six individual same-sign and opposite-sign channels, corresponding to an integrated luminosity of 2.1 fb$^{-1}$. A simple $\chi^2$ function is used to estimate the agreement of the model predictions with data, with the systematic and statistical errors on the background summed in quadrature. In particular, this allows us to find for each benchmark scenario the regions in the flavour parameters that are excluded at the 95\% confidence level (CL). 
Our results are summarised in Fig.~\ref{fig:lim}, in terms of the two independent mixing parameters $V_{eN}^R$, $V_{\mu N}^R$.
The red regions correspond to 95\% CL exclusion from ATLAS data, using the aforementioned procedure. The black outer arc $|V_{eN}^R|^2 + |V_{\mu N}^R|^2 = 1$ corresponds to no $\tau$ mixing, $V_{\tau N}^R = 0$, which includes the previously used benchmarks represented by small circles. Our plots clearly show that data allow to go beyond this assumption and explore scenarios with large $\tau$ mixing, since there is significant experimental sensitivity. 
These results should motivate dedicated experimental searches using more general benchmarks with non-zero $V_{\tau N}^R$.
In particular, we observe that heavy neutrinos with mixing $V_{\tau N}^R \leq 0.94$ are excluded for the second scenario. 
The range of mixing probed will extend farther when the full 5 fb$^{-1}$ dataset collected at 7 TeV is analysed. We have estimated the expected sensitivity by rescaling the backgrounds in Ref.~\cite{atlas} and making the assumption that the number of observed events equals the expected background. Remarkably, it is found that in the second benchmark the larger data sample would allow to exclude $V_{\tau N}^R \lesssim 0.987$. Eventually, if the entire range $V_{\tau N}^R \leq 1$ is excluded, that would only mean that the presence of heavy neutrinos is excluded, for this given set of masses $\mnmw$ and arbitrary flavour mixing.

We note that the simultaneous presence of non-zero $V_{eN}^R$, $V_{\mu N}^R$ leads to lepton-flavour-violating signals at low energy, so far unobserved. In both scenarios we have explicitly checked that in all the parameter space $|V_{eN}^R|^2 + |V_{\mu N}^R|^2 \leq 1$ the rate for $\mu \to e \gamma$~\cite{Cirigliano:2004mv} can be suppressed below the present upper bound, $\text{Br} (\mu \to e \gamma) < 2.4 \times 10^{-12}$ at the 90\% CL~\cite{Adam:2011ch}, by suitable choices of the masses of the two heavier neutrinos $m_{N_2},m_{N_3} \geq M_{W_R}$ and the third independent mixing angle in the $3\times 3$ unitary matrix of entries $V_{\ell N_i}^R$. (The contribution to $\tau \to e \gamma$ and $\tau \to \mu \gamma$ is well below present bounds.)
Interestingly, these considerations also highlight the synergy between low- and high-energy observables. While radiative decays, such as $\mu \to e \gamma$, are sensitive to higher scales than those to be directly probed at the LHC, they also sum all the three heavy neutrino contributions. On the other hand, the LHC has the potential to pinpoint their individual couplings, providing detailed information that is not accessible at low energies. In particular, when both $V_{eN}^R$ and $V_{\mu N}^R$ are non-zero the observation of lepton flavour violating (LFV) signals at the LHC, namely the production of $e\mu$ pairs without extra neutrinos, is posssible even if those effects are suppressed at low energies. The presence of these LFV signals could be spotted by an upper cut on missing energy, so as to remove the contribution from $\tau$ decays. However, this interesting issue is beyond the scope of the present work.

In summary, we have reconsidered the current LHC searches for heavy neutrinos in the context of left-right symmetric models. In contrast with earlier studies, which assume that heavy neutrinos only couple to $e$ and/or $\mu$, we have emphasised that a proper inclusion of flavour, namely arbitrary heavy neutrino mixing with the three charged leptons $e,\mu,\tau$, is possible. In particular, we have explicitly shown that current analyses~\cite{atlas} can be used to provide significant contraints on the general flavour mixing of heavy neutrinos. Our results, in view of the rich flavour structure observed in the light neutrino mixing, should motivate ``fully flavoured'' heavy neutrino analyses, beyond the simple benchmarks previously used.

{\it Acknowledgements.} We thank F. del \'Aguila for useful discussions. This work has been 
partially supported by MICINN projects
FPA2006-05294, FPA2010-17915, FPA2011-22975
and MULTIDARK CSD2009-00064 (Consolider-Ingenio 2010 Programme), by
Prometeo/2009/091 (Generalitat Valenciana), by the EU ITN UNILHC
PITN-GA-2009-237920 and Junta de Andaluc\'{\i}a projects FQM 101, FQM
03048 and FQM 6552. The work of O.K. has been supported by a CPAN fellowship.


\begin{thebibliography}{99}

\bibitem{Nakamura:2010zzi} 
  K.~Nakamura {\it et al.}  [Particle Data Group Collaboration],
  J.\ Phys.\ G G {\bf 37}, 075021 (2010).

\bibitem{Schwetz:2008er} 
  T.~Schwetz, M.~A.~Tortola and J.~W.~F.~Valle,
  New J.\ Phys.\  {\bf 10}, 113011 (2008).

\bibitem{Minkowski:1977sc} 
  P.~Minkowski,
  Phys.\ Lett.\ B {\bf 67}, 421 (1977).
\bibitem{Mohapatra:1979ia} 
  R.~N.~Mohapatra and G.~Senjanovic,
  Phys.\ Rev.\ Lett.\  {\bf 44}, 912 (1980).


\bibitem{Dittmar:1989yg} 
  M.~Dittmar, A.~Santamaria, M.~C.~Gonzalez-Garcia and J.~W.~F.~Valle,
  Nucl.\ Phys.\ B {\bf 332}, 1 (1990).

\bibitem{Buchmuller:1990vh} 
  W.~Buchmuller and C.~Greub,
  Phys.\ Lett.\ B {\bf 256}, 465 (1991).

\bibitem{Datta:1993nm} 
  A.~Datta, M.~Guchait and A.~Pilaftsis,
  Phys.\ Rev.\ D {\bf 50}, 3195 (1994).


\bibitem{Nardi:1994iv} 
  E.~Nardi, E.~Roulet and D.~Tommasini,
  Phys.\ Lett.\ B {\bf 327}, 319 (1994).

\bibitem{Tommasini:1995ii} 
  D.~Tommasini, G.~Barenboim, J.~Bernabeu and C.~Jarlskog,
  Nucl.\ Phys.\ B {\bf 444}, 451 (1995).


\bibitem{delAguila:2008pw} 
  F.~del Aguila, J.~de Blas and M.~Perez-Victoria,
  Phys.\ Rev.\ D {\bf 78}, 013010 (2008);
  F.~del Aguila, J.~A.~Aguilar-Saavedra, J.~de Blas and M.~Perez-Victoria,
  arXiv:0806.1023 [hep-ph];
  F.~del Aguila, J.~A.~Aguilar-Saavedra and J.~de Blas,
  PoS ICHEP {\bf 2010}, 296 (2010).

\bibitem{del Aguila:2006dx} 
  F.~del Aguila, J.~A.~Aguilar-Saavedra and R.~Pittau,
  J.\ Phys.\ Conf.\ Ser.\  {\bf 53}, 506 (2006).

\bibitem{delAguila:2009bb} 
  F.~del Aguila, J.~A.~Aguilar-Saavedra and J.~de Blas,
  Acta Phys.\ Polon.\ B {\bf 40}, 2901 (2009).


\bibitem{Keung:1983uu} 
  W.~-Y.~Keung and G.~Senjanovic,
  Phys.\ Rev.\ Lett.\  {\bf 50}, 1427 (1983).

\bibitem{delAguila:2007em} 
  F.~del Aguila, J.~A.~Aguilar-Saavedra and R.~Pittau,
  JHEP {\bf 0710}, 047 (2007).

\bibitem{delAguila:2007ua} 
  F.~del Aguila and J.~A.~Aguilar-Saavedra,
  JHEP {\bf 0711}, 072 (2007).

\bibitem{AguilarSaavedra:2009ik} 
  J.~A.~Aguilar-Saavedra,
  Nucl.\ Phys.\ B {\bf 828}, 289 (2010).


\bibitem{Ferrari:2000sp} 
  A.~Ferrari, J.~Collot, M-L.~Andrieux, B.~Belhorma, P.~de Saintignon, J-Y.~Hostachy, P.~.Martin and M.~Wielers,
  Phys.\ Rev.\ D {\bf 62}, 013001 (2000).

\bibitem{Gninenko:2006br} 
  S.~N.~Gninenko, M.~M.~Kirsanov, N.~V.~Krasnikov and V.~A.~Matveev,
  Phys.\ Atom.\ Nucl.\  {\bf 70}, 441 (2007).

\bibitem{CMSN}
CMS Collaboration, note CMS PAS EXO-11-002.

\bibitem{Nemevsek:2011hz} 
  M.~Nemevsek, F.~Nesti, G.~Senjanovic and Y.~Zhang,
  Phys.\ Rev.\ D {\bf 83}, 115014 (2011).

\bibitem{atlas}
  G.~Aad {\it et al.}  [ATLAS Collaboration],
  arXiv:1203.5420 [hep-ex].

\bibitem{An:2012eh} 
  F.~P.~An {\it et al.}  [DAYA-BAY Collaboration],
  arXiv:1203.1669 [hep-ex].


\bibitem{del Aguila:2005mf} 
  F.~del Aguila, J.~A.~Aguilar-Saavedra, A.~Martinez de la Ossa and D.~Meloni,
  Phys.\ Lett.\ B {\bf 613}, 170 (2005);
  F.~del Aguila and J.~A.~Aguilar-Saavedra,
  JHEP {\bf 0505}, 026 (2005).

\bibitem{Schechter:1980gr} 
  J.~Schechter and J.~W.~F.~Valle,
  Phys.\ Rev.\ D {\bf 22}, 2227 (1980).

\bibitem{Gavin:2010az} 
  R.~Gavin, Y.~Li, F.~Petriello and S.~Quackenbush,
  Comput.\ Phys.\ Commun.\  {\bf 182}, 2388 (2011).

\bibitem{delAguila:2008cj} 
  F.~del Aguila and J.~A.~Aguilar-Saavedra,
  Nucl.\ Phys.\ B {\bf 813}, 22 (2009);
  Phys.\ Lett.\ B {\bf 672}, 158 (2009).

\bibitem{Sjostrand:2006za}
  T.~Sjostrand, S.~Mrenna and P.~Z.~Skands,
  JHEP {\bf 0605}, 026 (2006).

\bibitem{RichterWas:2002ch}
  E.~Richter-Was,
  arXiv:hep-ph/0207355.

\bibitem{Barenboim:1996nd} 
  G.~Barenboim, J.~Bernabeu, J.~Prades and M.~Raidal,
  Phys.\ Rev.\ D {\bf 55}, 4213 (1997);
  A.~Maiezza, M.~Nemevsek, F.~Nesti and G.~Senjanovic,
  Phys.\ Rev.\ D {\bf 82}, 055022 (2010).

\bibitem{Cirigliano:2004mv} 
  V.~Cirigliano, A.~Kurylov, M.~J.~Ramsey-Musolf and P.~Vogel,
  Phys.\ Rev.\ D {\bf 70}, 075007 (2004).

\bibitem{Adam:2011ch} 
  J.~Adam {\it et al.}  [MEG Collaboration],
  Phys.\ Rev.\ Lett.\  {\bf 107}, 171801 (2011).


\end{thebibliography}
\end{document}